\documentstyle[psfig]{mn}

\newif\ifAMStwofonts

\def\bjr{$\rm B_J - R$}
\def\bj{$\rm B_J$}
\def\ie{\protect\rm i.e.\ }
\def\etal{et~al.}

\def\lya{Ly$\alpha$}
\def\simgt{$_>\atop{^\sim}$}
\def\simlt{$_<\atop{^\sim}$}

\ifoldfss
  \ifCUPmtlplainloaded \else
    \NewTextAlphabet{textbfit} {cmbxti10} {}
    \NewTextAlphabet{textbfss} {cmssbx10} {}
    \NewMathAlphabet{mathbfit} {cmbxti10} {} 
    \NewMathAlphabet{mathbfss} {cmssbx10} {} 
  \fi
  \ifAMStwofonts
    \ifCUPmtlplainloaded \else
      \NewSymbolFont{upmath} {eurm10}
      \NewSymbolFont{AMSa} {msam10}
      \NewMathSymbol{\upi}     {0}{upmath}{19}
      \NewMathSymbol{\umu}     {0}{upmath}{16}
      \NewMathSymbol{\upartial}{0}{upmath}{40}
      \NewMathSymbol{\leqslant}{3}{AMSa}{36}
      \NewMathSymbol{\geqslant}{3}{AMSa}{3E}

       \let\le=\leqslant
      \let\geq=\geqslant 
    \fi
  \fi
\fi 

\ifnfssone
  \newmathalphabet{\mathit}
  \addtoversion{normal}{\mathit}{cmr}{m}{it}
  \addtoversion{bold}{\mathit}{cmr}{bx}{it}
  \newmathalphabet{\mathbfit} 
  \addtoversion{normal}{\mathbfit}{cmr}{bx}{it}
  \addtoversion{bold}{\mathbfit}{cmr}{bx}{it}
  \newmathalphabet{\mathbfss} 
  \addtoversion{normal}{\mathbfss}{cmss}{bx}{n}
  \addtoversion{bold}{\mathbfss}{cmss}{bx}{n}
  \ifAMStwofonts
    \ifCUPmtlplainloaded \else
      \UseAMStwoboldmath
      \makeatletter
      \new@mathgroup\upmath@group
      \define@mathgroup\mv@normal\upmath@group{eur}{m}{n}
      \define@mathgroup\mv@bold\upmath@group{eur}{b}{n}
      \edef\UPM{\hexnumber\upmath@group}
      \new@mathgroup\amsa@group
      \define@mathgroup\mv@normal\amsa@group{msa}{m}{n}
      \define@mathgroup\mv@bold\amsa@group{msa}{m}{n}
      \edef\AMSa{\hexnumber\amsa@group}
      \makeatother
      \mathchardef\upi="0\UPM19
      \mathchardef\umu="0\UPM16
      \mathchardef\upartial="0\UPM40
      \mathchardef\leqslant="3\AMSa36
      \mathchardef\geqslant="3\AMSa3E

       \let\le=\leqslant
      \let\geq=\geqslant 
    \fi
  \fi
\fi 

\ifnfsstwo
  \DeclareMathAlphabet{\mathbfit}{OT1}{cmr}{bx}{it}
  \SetMathAlphabet\mathbfit{bold}{OT1}{cmr}{bx}{it}
  \DeclareMathAlphabet{\mathbfss}{OT1}{cmss}{bx}{n}
  \SetMathAlphabet\mathbfss{bold}{OT1}{cmss}{bx}{n}
  \ifAMStwofonts
    \ifCUPmtlplainloaded \else
      \DeclareSymbolFont{UPM}{U}{eur}{m}{n}
      \SetSymbolFont{UPM}{bold}{U}{eur}{b}{n}
      \DeclareSymbolFont{AMSa}{U}{msa}{m}{n}
      \DeclareMathSymbol{\upi}{0}{UPM}{"19}
      \DeclareMathSymbol{\umu}{0}{UPM}{"16}
      \DeclareMathSymbol{\upartial}{0}{UPM}{"40}
      \DeclareMathSymbol{\leqslant}{3}{AMSa}{"36}
      \DeclareMathSymbol{\geqslant}{3}{AMSa}{"3E}

       \let\le=\leqslant
      \let\geq=\geqslant 
    \fi
  \fi
\fi 

\ifCUPmtlplainloaded \else
  \ifAMStwofonts \else 
    \def\upi{\pi}
    \def\umu{\mu}
    \def\upartial{\partial}
  \fi
\fi

\title{The Second APM UKST Colour Survey for z$>$4 Quasars}

\author[L. J. Storrie-Lombardi et al.]
{Lisa J. Storrie-Lombardi$^1$, Michael J. Irwin$^2$, Richard G. McMahon$^2$
\newauthor and Isobel M. Hook$^3$\\
$^1$ SIRTF Science Center, California Institute of Technology, MS 100-22, Pasadena, CA 91125, USA \\
$^2$ Institute of Astronomy, Madingley Road, Cambridge, CB3 0HA, England, UK\\ 
$^3$ Institute for Astronomy, Royal Observatory, Blackford Hill, 
Edinburgh, EH9 3HJ, Scotland, UK} 

\date{Accepted version 1 December 2000 }

\pagerange{\pageref{firstpage}--14}
\pubyear{2000}

\begin{document}

\maketitle

\label{firstpage}

\begin{abstract}
We present the spectra, positions, and finding charts for 31 bright
(R $<$ 19.3) colour-selected quasars covering the redshift range
z = 3.85 -- 4.78, with 4 having redshifts z $>$ 4.5. 
The majority are in the southern sky ($\delta < -25^\circ$). 
The quasar candidates were selected for their red ({\bjr} \simgt 2.5) colours
from UK or POSSII Schmidt Plates scanned at the Automated Plate Measuring
facility in Cambridge. Low resolution (\simgt 10 \AA) spectra were obtained to
identify the quasars, primarily at the Las Campanas Observatory.
The highest redshift quasar in our survey is at z $\approx$ 4.8
(R = 18.7) and its spectrum shows a damped 
{\lya} absorption system at z = 4.46.
This is currently the highest redshift damped {\lya} absorber detected.
Five of these quasars exhibit intrinsic broad absorption line features.
Combined with the previously published results from the first
part of the APM UKST survey
we have now surveyed a total of $\sim$8000deg$^2$ of sky i.e. 40\% of
the high galactic latitude($|b|>30^\circ$) sky, resulting 
in 59 optically selected quasars in the redshift range 3.85 to 4.78; 
49 of which have z$\geq$4.00. 

\end{abstract}


\section{Introduction}

High redshift quasars provide a powerful means for exploring early epochs.
It is likely that they flag regions where galaxy formation is very active.  
Their host galaxies are probably still forming and they may occur 
in the exceptional `5 $\sigma$' peaks in the matter distribution of 
the early Universe.
In addition to being of intrinsic interest themselves,
bright high redshift quasars are particularly valuable as probes of
the intervening gas clouds and galaxies superimposed on
their spectra in absorption.  The galaxies that intercept
their line-of-sight provide samples selected by gas
cross-section, without regard to their surface brightness,
luminosity, or star formation rate.
Though direct studies of high redshift galaxies are now possible,
those selected by the absorption lines they produce in quasar spectra
still provide the only means to study in detail their kinematic
properties at high resolution. This information can be combined
with the colour and morphological information obtained from imaging
to provide a complete picture of individual galaxies at high redshift.

In 1996 we published spectra for 28 quasars discovered in the first
APM Colour Survey for high redshift quasars (Storrie-Lombardi \etal\ 1996, hereafter APM1).
These were located at equatorial declinations.
We have completed the second APM Colour Survey for bright, z $>$ 4 quasars
in the southern hemisphere (mainly with $\delta < -25$), discovering 23 more 
high redshift quasars.  We also include 8 previously
unpublished quasars also found using this same technique that were 
not directly part of the Las Campanas follow-up campaign.   
Three other lower redshift quasars (z = 0.40 -- 2.73) were also found 
serendipitously in the survey and we include their spectra for completeness.

The paper is organised as follows. 
In $\S$2 we discuss the methodology of the quasar candidate selection,
in $\S$3 we describe the observations and present the spectra, in $\S$4 
we discuss some of the objects individually, and in  $\S$5 we provide 
a summary discussion. 

\section{Quasar Candidate Selection}

The quasar candidates were generally selected of the basis of their exceptionally
red {\bjr} colour (Irwin, McMahon \& Hazard 1991), although in a few 
cases the addition of an I passband enabled selection closer to the main 
stellar locus to be made.  All of the quasars bar one, BR J1603$+$0721, were
found using \bj, R\footnote{For simplicity we will use R to denote either R or OR (the majority)
plates.  The OR passband is the official UKST survey band and covers the
wavelength range 5900\AA\ -- 6900\AA.  Earlier `B' grade survey plates were 
often taken in the R passband covering 6300\AA\ -- 6900\AA, some of which were
used during the course of our survey.}
and, occasionally, I passband UK Schmidt plates.
In all cases these plates came from the generic southern sky survey material
taken by the UK Schmidt telescope (UKST) and were either glass copies of `A' 
grade survey {\bj} plates, or original survey OR/R and I plates.   
The Northern object was found as part of a test series of measurements of 
POSSII survey (\ie second epoch Palomar Sky Survey) {\bj} glass copies 
matched to POSSII R survey film copies.  

All plate material was measured and analysed at the Automated Plate Measuring 
(APM) facility in Cambridge, UK, to produce image lists including 
classification, magnitude and colour information (for further details see
Kibblewhite {\etal} 1984; Irwin, Demers \& Kunkel 1990).  A large fraction
of the UKST survey plates were measured over a 5 year period from 1989.
Consequently, many of the final survey grade UKST plates were not available
at scanning time and in several cases early `B' grade survey plates were used 
instead.

At an early stage in the programme we decided to restrict the candidate
selection to those stellar objects lying well away from the main stellar
locus and to relatively bright magnitudes.   This was mainly to increase the 
efficiency of the spectrographic follow-up and also because the primary goal of
the programme was to find a bright sample of quasars for further absorption 
line follow-up studies.  In addition, the multi-epoch nature of the plate 
material precludes attaining completeness based solely on colour/magnitude 
information due to the intrinsic variability of quasars (see for example 
Hook \etal\ 1994 and references therein).  The effects of colour selection on
sample completeness have been thoroughly investigated over the past few years 
(eg. Warren, Hewett \& Osmer 1994; APM1, Kennefick \etal\ 1995 and references 
therein).

\begin{figure}
\centerline{\psfig{figure=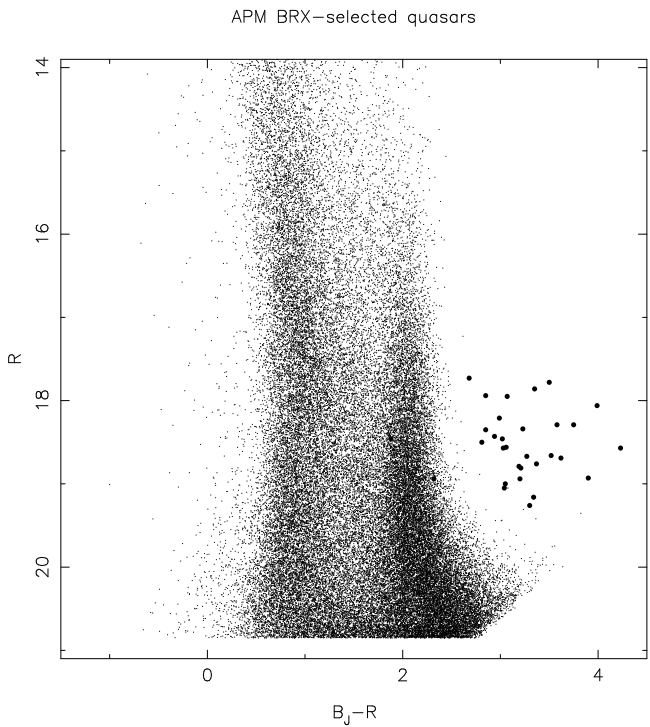,height=12.0cm}}
\caption{A \bj,R
colour-magnitude diagram for a typical high latitude UKST field used 
in the APM survey.  Every
detected B$_{\rm J}$,R matched pair of objects classified as stellar on the R
plate is plotted as small dot.  Overlaid as filled circles are the complete
southern sample of BRX-selected quasars.}
\label{f_cmd}
\end{figure}

The efficacy of the traditional two-colour selection for finding z $>$ 4 
quasars based on \bj,R,I photometry is shown in fig. 1 of Irwin, 
McMahon \& Hazard (1991).  Since most of the current sample of quasars were 
selected using what we have called the `BRX' technique, we 
demonstrate this method in the current paper.  Fig.~\ref{f_cmd} shows a \bj,R
colour-magnitude diagram for a typical high latitude UKST field.  Every
detected \bj,R matched pair of objects classified as stellar on the R 
plate is plotted as a small dot.  Overlaid as filled circles are the complete 
southern sample of BRX-selected quasars.  Of the roughly 250,000 paired objects
on each high latitude UKST field, two-thirds are classified as stellar on the R plate
and roughly 50,000 of these are brighter than R $=$ 19 -- 19.5, the range for 
the R magnitude limit.  Although the aim was to find bright R \simlt\ 19
magnitude quasars, rather than impose a rigid magnitude cut we allowed the
faint limit to reach 19.5 if the plate pairs were of suitably good quality
and had a clean colour-magnitude diagram at this limit.  

The red boundary for BR candidate selection was set to approximately
\bj--R $=$ 2.5 for images brighter than R $=$ 18.5 and was then increased
roughly linearly to \bj--R $=$ 3 at R $=$ 19.5.  This results in a very 
clean sample, as can be seen from fig.~\ref{f_cmd}, with usually at most 10 
candidates per field.  Roughly half of the candidates can be easily rejected
using the online APM catalogue finding charts.  These rejected objects would 
typically be objects: close to the edge of the scanned area on one or other
plate; in, or near, the Halo of bright stars; affected by some scratch or 
satellite trail; close to one of the density wedges; and so on.

\begin{figure*}
\begin{minipage}{115mm}
\centerline{\psfig{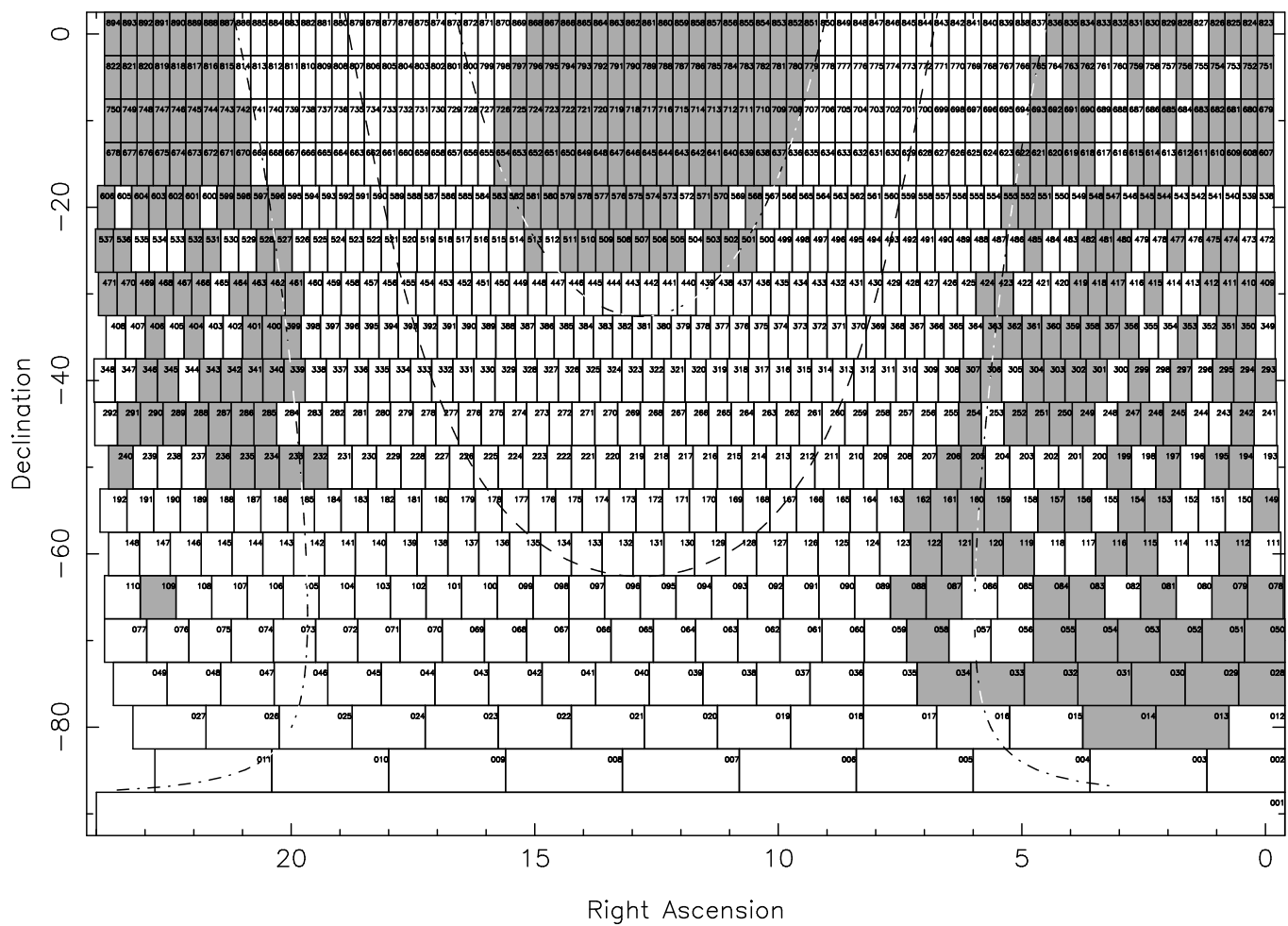}}
\caption{Plotted here is a schematic representation of the area of sky surveyed in the
current work, together with those areas surveyed in our previously published sample
(Storrie-Lombardi \etal\ 1996).   The total area of Southern high latitude sky surveyed is
roughly 8000 sq deg from a total of 328 UKST fields.  The dot-dash lines denote
the $|b| = 30^\circ$ Galactic latitude locii.  }
\label{f_qsostatus}
\end{minipage}
\end{figure*}

A schematic representation of the area of sky surveyed in the current work,
together with those areas surveyed in our previously published sample 
(APM1), is given in fig.~\ref{f_qsostatus}.   The total area of 
Southern high latitude sky surveyed is roughly 8000 square degrees from a total
of 328 UKST fields.  Although the measured area of each field amounts to some
5.8 $\times$ 5.8 degrees, the effective area of each field is only 
$\approx$25 square degrees.  This is mainly due to the 5 degree grid spacing of the 
survey but also involves reductions because of the effect of density wedges
and  general edge effects.  The majority of the fields used are high 
latitude in the sense that $|b| > 30^\circ$, however there are a small number 
of fields closer to the Galactic Plane than that, since the survey data was 
measured for a variety of projects.
 
External deep photometric calibration did not (and still does not) exist for
the majority of the fields surveyed.   However, as in earlier work using
the APM facility, we used an internal plate calibration method (Bunclark \& 
Irwin 1983) as the basis of a (mainly) internal calibration scheme.  Making
use of the uniformity in depth of the Schmidt survey plates and the fact that 
colour equations for UKST plate/filter combinations are well known 
(eg. Blair \& Gilmore 1982; Irwin, Demers \& Kunkel 1990) facilitates use of 
a natural photographic passband magnitude scale.   Although from external
sequence checks the faint stellar R-band magnitude scale can only be tied down
to an $rms$ variation of 0.25 magnitudes with this method, the known 
properties of Galactic foreground stars can be readily used to define the 
{\bjr} colour to an accuracy of 0.1 magnitude.  This, and a reliable star--galaxy
classifier, make candidate selection extremely straightforward.

\section{Observations}

Candidate follow-up proceeded over several years, using a range of facilities 
(see Table~\ref{t1} column 10), with the majority of the follow-up taking 
place at Las Campanas during three spectroscopic runs in 1997 and 1998.
Candidates were generally prioritised for observation on the extremity
of their {\bjr} colour and on their relative brightness.  Although we have not
observed more than $\approx$50\% of the total number of candidates, spectra 
have been obtained for all of
the most promising objects.  In several fields where
early on during the survey we observed all the candidates, we found that 
objects near the stellar locus invariably turned out to be late M-stars
shifted out of the main locus by the inevitable non-Gaussian tale of the
photometric errors.  In subsequent observing runs we were more conservative
in candidate selection in order to speed up the progress of the survey.

For the follow-up undertaken at the Dupont 100-inch telescope at Las Campanas 
we used the Modular Spectrograph with the 300 l/mm grating and 
the Tek\#5 CCD, in gray and bright time.  This gives wavelength coverage from 
approximately 4000--9000\AA\ (2.5\AA \ per pixel) 
which easily covers the {\lya} and CIV emission 
lines and the {\lya} forest drop for quasars with 4 $<$ z $<$ 5. 
Typical exposure times were 600 -- 900 seconds which allowed differentiation
between high redshift quasars with a continuum drop across the {\lya}
line caused by intervening cosmological absorption, and M-dwarfs or 
galaxies at z $\sim$ 0.3, the main contaminants in the survey. The spectra
were reduced as they were taken using standard IRAF\footnote{IRAF is distributed by the
National Optical Astronomy Observatories, which is operated by the
Association of Universities for Research in Astronomy, Inc.~(AURA)
under cooperative agreement with the National Science Foundation.}
routines which allowed us
to follow-up in real time any candidates where the first spectrum did not make
it immediately clear whether it was a quasar or not.
For every 
8--10 candidates remaining from the selection in $\S$2, one is a high redshift
quasar.  The additional observations were also taken at low resolution on a 
variety of facilities during other observing runs between 1986-1997, mainly 
at times when the primary observing programme could not be executed. 

Throughout the survey, apart from high redshift quasars, the only `real' 
objects we have found lying significantly redward of the K/M stellar locus are:
misclassified compact galaxies, usually ellipticals at redshift 0.3--0.4, 
where the 4000\AA \ break has left the \bj\ passband at 5400\AA; 
very late-type M giants including Miras and other long period variables; 
distant Halo carbon stars (eg. Totten \& Irwin 1998); 
very late type - often high proper motion - nearby dwarf
M stars (Irwin, McMahon \& Reid 1991; Kirkpatrick, Todd \& Irwin 1997) and at 
least one field brown dwarf (Tinney 1998); and the occasional CV and/or PN.

The quasars discovered are listed in Table~\ref{t1}.  The quasars were 
originally selected off photographic plates using B1950 coordinates as the 
default equinox, but we have listed their names with the J2000 coordinate 
system as well for ease of cross-reference. 
Columns 1, 2, and 3 list the quasar name, right  
ascension and declination in B1950 coordinates and columns 4,5, and 6 give the 
same information in the J2000 equinox. 
Columns 7, 8, and 9 list the APM R magnitude, the APM {\bjr} colour, and the plate 
number off which the object information was measured.
Column 10 lists where and when the observations were made.  The abbreviations
are:  LCO = Dupont 100-inch -- Las Campanas Observatory,  AAT = 4.3-m Anglo Australian Telescope, 
WHT = 4.2-m William Herschel Telescope, CTIO = Blanco 4-m -- Cerro Tololo Inter-American Observatory, 
and KPNO = Mayall 4-m -- Kitt Peak National Observatory.  Column 11 gives the 
total exposure time for each spectrum, and column 12 the quasar redshifts determined
from these spectra. The redshifts were generally determined from the blueward edge 
of the {\lya} emission unless the spectrum had a high enough signal-to-noise ratio 
to measure the CIV 
emission line.  Previous experience has shown that the uncertainties in the redshifts
measured from these discovery spectra will be $\pm0.1$.
Five of the quasars exhibit intrinsic broad absorption line (BAL) features.
These are noted in the table.
The spectra are shown in fig.~\ref{f_spectra} and the finding charts in fig.~\ref{f_fc}.
The O$_2$ A-band absorption feature at 7600\AA\ has not been removed from any of the spectra.
In addition to the 31 high redshift quasars we also list 3 additional lower redshift
objects discovered as part of our
Las Campanas survey. These include a broad absorption line quasar at z = 2.73 and 
quasars at redshifts z = 0.40 and z = 0.68. 

\section{Notes on Individual Objects}

Higher resolution spectroscopy of these quasars 
is necessary to do quantitative studies of their emission and 
absorption line properties (see P\'eroux \etal\ 2001) but
many interesting features are apparent in the discovery spectra.

\smallskip
\noindent (1) BR J0004-6655, z = 2.73 BAL

This quasar exhibits broad absorption lines.  It is intrinsically red 
in the optical {\bjr} colours due to the broad absorption
line troughs redward of 5900\AA.

\noindent (2) BR J0006-6208, z = 4.51

This spectrum shows evidence for two damped {\lya} absorption candidates
at z $\approx$ 3.2 and z $\approx$ 3.8 and a Lyman limit system at z $\approx$ 3.2.  The {\lya} emission line is relatively weak, but not atypical
of z $>$ 4 quasars.

\noindent (3) BR J0018-3527, z = 4.15 BAL

This quasar exhibits broad absorption line features.

\noindent (4) BR J0030-5129, z = 4.17

This quasar shows very strong, peaky emission lines and in conjunction with the
previous two objects demonstrates the wide variety of quasar spectra found in
photographic multicolour surveys.

\noindent (5) BR J0046-1606, z = 3.85 BAL

This quasar exhibits broad absorption line features which enhance the red optical
colour at this relatively low redshift.

\noindent (6) BRI J0048-2442, z = 4.15 \\
\noindent (7) BRI J0113-2803, z = 4.30 \\
\noindent (8) BRI J0137-4224, z = 3.97

Nothing much is evident in these three $\approx$ 50\AA\ resolution spectra 
taken in the late 1980s, other than the fact that these objects are quasars.  
Higher resolution spectra have been taken of these quasars in a recent survey 
for absorption lines systems (Storrie-Lombardi \& Wolfe 2000).  BRI J0137-4224 
is of interest since it was the first APM BRI-selected quasar to be found, 
thereby proving the concept, and was discovered at CTIO in 1986 shortly after 
the first redshift 4 quasar was found by Warren \etal\ (1987).

\noindent (9) BR J0234-1806, z = 4.30

This quasar has strong {\lya} emission.  The apparently negative flux regions
around 4000--5000\AA\ are due to a combination of poor signal-to-noise
and imperfect sky subtraction.

\noindent (10) BR J0301-5537, z = 4.11

A well-defined Lyman limit system is apparent at z $\approx$ 4.0. 

\noindent (11) BR J0302-0156,  z = 4.25 BAL

This quasar shows broad absorption line features. 

\noindent (12) BR J0307-4945,  z = 4.78

This is the highest redshift quasar in our survey.  It shows a damped
{\lya} absorption feature at z = 4.46.  Both the {\lya} and Ly$\beta$ lines 
are visible at 6650\AA\ and 5605\AA.  This absorption system is discussed in 
more detail in McMahon \etal\ (2001), P\'eroux \etal\ 2001, and
Dessauges-Zavadsky \etal\ (2001).  It is currently 
the highest redshift damped absorber known.

\noindent (13) BR J0311-1727, z = 4.00

A candidate damped {\lya} absorber is detected at z $\approx$ 3.7.

\noindent (14) BR J0324-2918, z = 4.62

This is the second highest redshift quasar in the current sample.

\noindent (15) BR J0334-1612, z = 4.32 

A higher resolution spectrum of this quasar is shown  
in Storrie-Lombardi \& Wolfe (2000). 

\noindent (16) BR J0355-3811, z = 4.58

An strong MgII absorption feature is evident at z =  1.99.

\noindent (17) BR J0415-4357, z = 4.08 

This is another quasar with a very strong {\lya} emission line.

\noindent (18) BR J0419-5716, z = 4.37 

No comments.

\noindent (19) BR J0426-2202, z = 4.30

This is a very poor signal-to-noise but not atypical discovery spectrum,
confirmed by later better quality spectroscopy.

\noindent (20) PMN J0525-3343, z = 4.40

This quasar was detected using the BRX technique and also independently discovered
as a radio-loud quasar by Hook \etal\ (2001).  The spectrum is unusual for
high redshift radio loud quasars as the {\lya} line is much weaker than
normally found.

\noindent (21) BR J0529-3526, z = 4.41
\noindent (22) BR J0529-3552, z = 4.15 

No comments.

\noindent (23) BR J0714-6455, z = 4.47 

This quasar shows a Lyman limit system (and possible damped {\lya} candidate) 
at z $\approx$ 4.4.  

\noindent (24) BR J1310-1740, z = 4.20 \\
\noindent (25) BR J1330-2522, z = 3.91

No comments.

\noindent (26) BR J1447-2117, z = 0.40

This low redshift quasar is relatively blue in {\bjr} and must have entered 
the sample through intrinsic variability.  The epoch difference between the
plate pairs was large (18 years).

\noindent (27) BR J1603+0721, z = 4.35

A higher resolution spectrum of quasar this quasar is  
shown in Storrie-Lombardi \& Wolfe (2000). 
This object is notable because it was discovered using the POSSII glass
and film copies.

\noindent (28) BR J2015-4032, z = 0.68

This low redshift quasar is relatively blue in {\bjr} and must have entered 
the sample through intrinsic variability.  The epoch difference between the
plate pairs was large (15 years). 

\noindent (29) BR J2017-4019, z = 4.15  BAL

The {\lya} and CIV emission lines in this quasar are almost completely absorbed
giving the spectrum the appearance of a step-function.

\noindent (30) BR J2131-4429, z = 3.83  BAL

This quasar shows classic broad absorption line features.

\noindent (31) BR J2216-6714, z = 4.49 

There is a possible double damped {\lya} absorber at z $\approx$ 4.3.

\noindent (32) BR J2317-4345, z = 4.02 

There is a damped {\lya} candidate at z = 3.4.

\noindent (33) BR J2328-4513, z = 4.38  \\
\noindent (34) BR J2349-3712, z = 4.21 

No comments.

\section{Discussion}

The only other comparable bright large area high redshift survey is that of
Kennefick, Djorgovski \& de Carvalho (1995 and references therein, see also
http://astro.caltech.edu/$\sim$george/z4.qsos). Although this survey is based on 
POSSII photographic plates B,R and I plates, the underlying methodology is 
essentially the same as that described in Irwin, McMahon \& Hazard (1991).  
In the first phase of this survey 10 quasars at redshifts $>$ 4 were found, 
selected from 27 fields covering an area of 681 deg$^2$.  

Combining the results from the First APM Colour Survey for High Redshift 
Quasars (APM1) with the work presented in this paper we have found a total of 
59 bright high redshift quasars (R $\le$ 19.5; 3.8 $<$ z $<$ 4.8) in a survey 
of approximately 8000 square degrees of the southern and equatorial sky. 
In fig.~\ref{f_zhist} we show two histograms of the redshift distribution of 
the combined APM Colour Surveys.  The left panel shows the combined histogram 
with the BRX-selected quasars shown with single hatch marks and the BRI-selected 
quasars shown with the double hatch marks.  The right panel again shows the 
combined survey histogram, with hatch marks overlaid on the quasars
that exhibit broad absorption lines (BAL) characteristics. 

The histograms highlight several important characteristics of the BR(I) survey:

a.) There is a well defined upper redshift limit to which the survey is
sensitive.  This upper limit is primarily caused by the R-band emulsion cutoff 
at 6900\AA \ and represents the point where the redshifted {\lya} line 
moves out of the R-band.  Consequently the entire R-band flux lies within the 
{\lya} forest and the intrinsic strong `continuum drop' across
{\lya} causes the R-band magnitude selection boundary to move to even 
brighter absolute magnitudes on the quasar luminosity function, with a 
corresponding dramatic fall in the expected number of quasars. 
The other compounding factor 
at z $>$ 5 is the statistical proximity
of Lyman limit systems to the redshift of the quasar (eg. Storrie-Lombardi 
\etal\ 1994). This means that even bright quasars are not detected on the {\bj}
plates, which have a limit of roughly {\bj} $= 22.5$.

b.) The roll-over in numbers at redshift z $=$ 4.2 is mainly caused by the 
candidate selection for the majority of the fields making use of the BRX 
selection technique.  Although, we are dealing with small number statistics, 
it is clear that the BRI technique can be used to somewhat lower redshifts, 
z $=$ 4.0, in general.  This is simply because the extra leverage obtained from 
the I-band enables candidates to be selected closer to the stellar locus.
For example, it is clear from comparing fig. 1. of Irwin, McMahon, \& Hazard, 
(1991), with figure 1. in the current paper, that BRX-selected candidates
are a subset of BRI-selected objects.  

c.) In general the BAL quasars appear to be at redshifts significantly lower
than ``normal'' quasars.  This is also a selection bias due to the fact that
in addition to the usual strong absorption troughs blueward of the standard
quasar emission lines, BAL quasars have strong Lyman limit systems at the
redshift of the quasar.  This depresses the {\bj} band flux more than
for normal quasars enabling them to be selected just below redshift z $=$ 4
and also causes the flux to be depressed to such a low level that they no
longer register on the {\bj} plates at redshifts much beyond z $=$ 4.3.
Indeed the highest redshift quasar without a measurable {\bj} flux
is a BAL quasar.  However, for most of the sample, we did not pursue candidates
not detected on the {\bj} plate.

\section*{Acknowledgements}

We thank the support staff at the Las Campanas, Cerro Tololo, Anglo-Australian,
Kitt Peak, and Royal Greenwich Observatories for their assistance in obtaining
these observations.  We thank the UKSTU for providing the plate material and 
thank the members of the APM facility, past and present, for maintaining such
an excellent system.

\bsp

\begin{table}
\caption{Second APM Colour Survey Quasars -- Journal of Observations }
\label{t1}
\end{table}

\clearpage
\begin{figure*}
\begin{minipage}{115mm}
\centerline{\psfig{figure=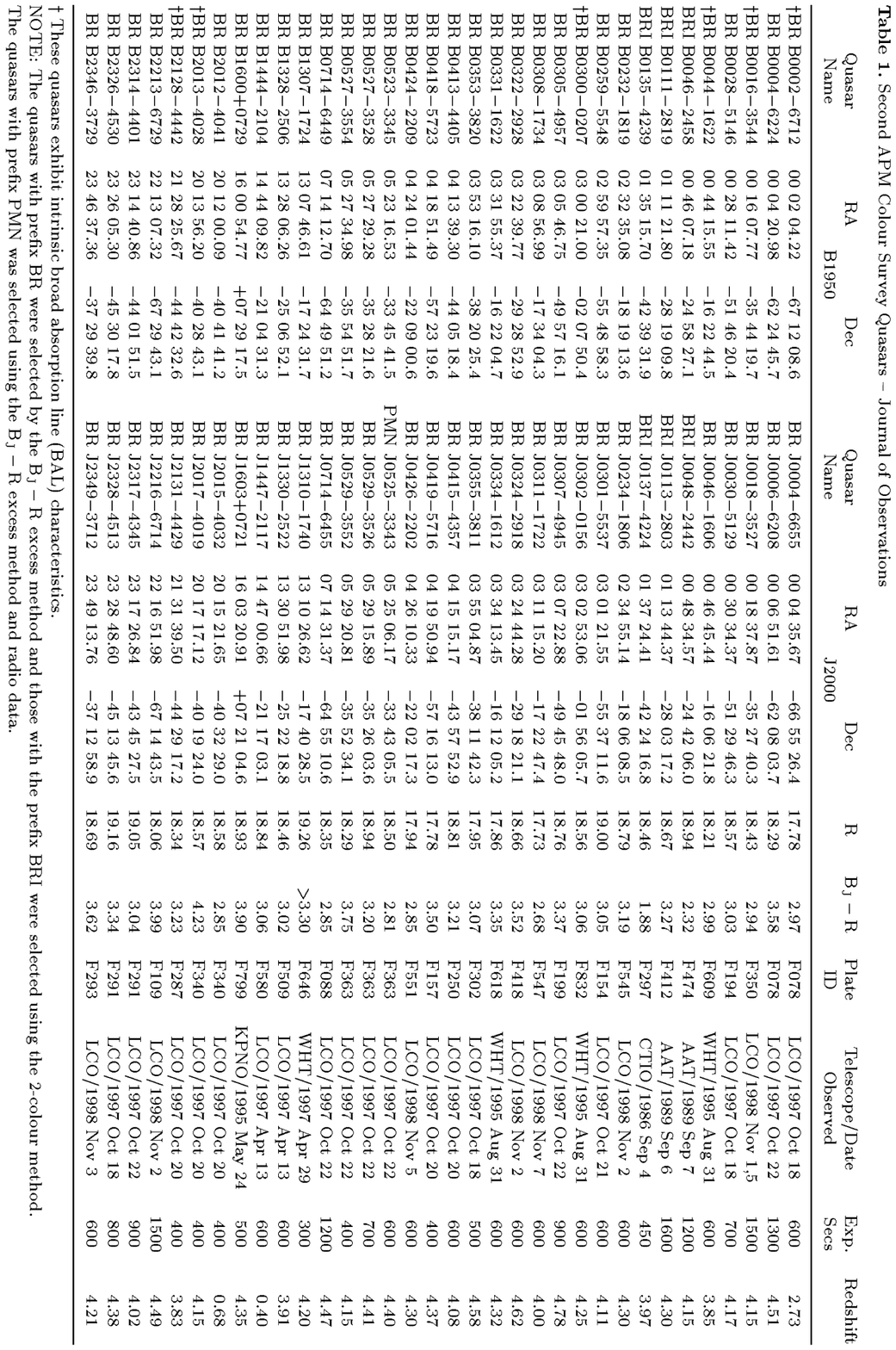,width=7.0in}}
\end{minipage}
\end{figure*}

\clearpage


\clearpage
\begin{figure}
\centerline{\psfig{figure=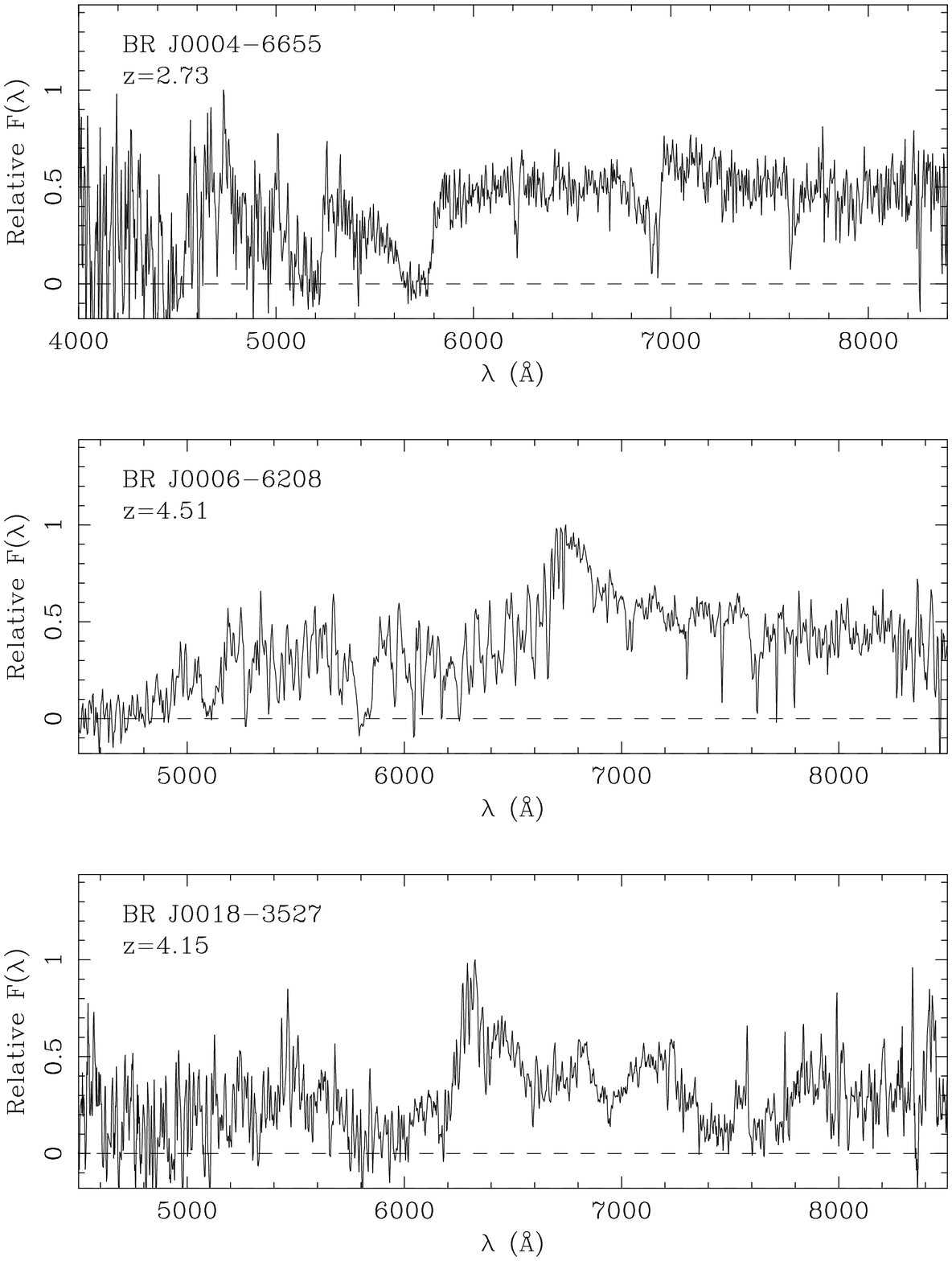,height=10.0cm}}
\centerline{\psfig{figure=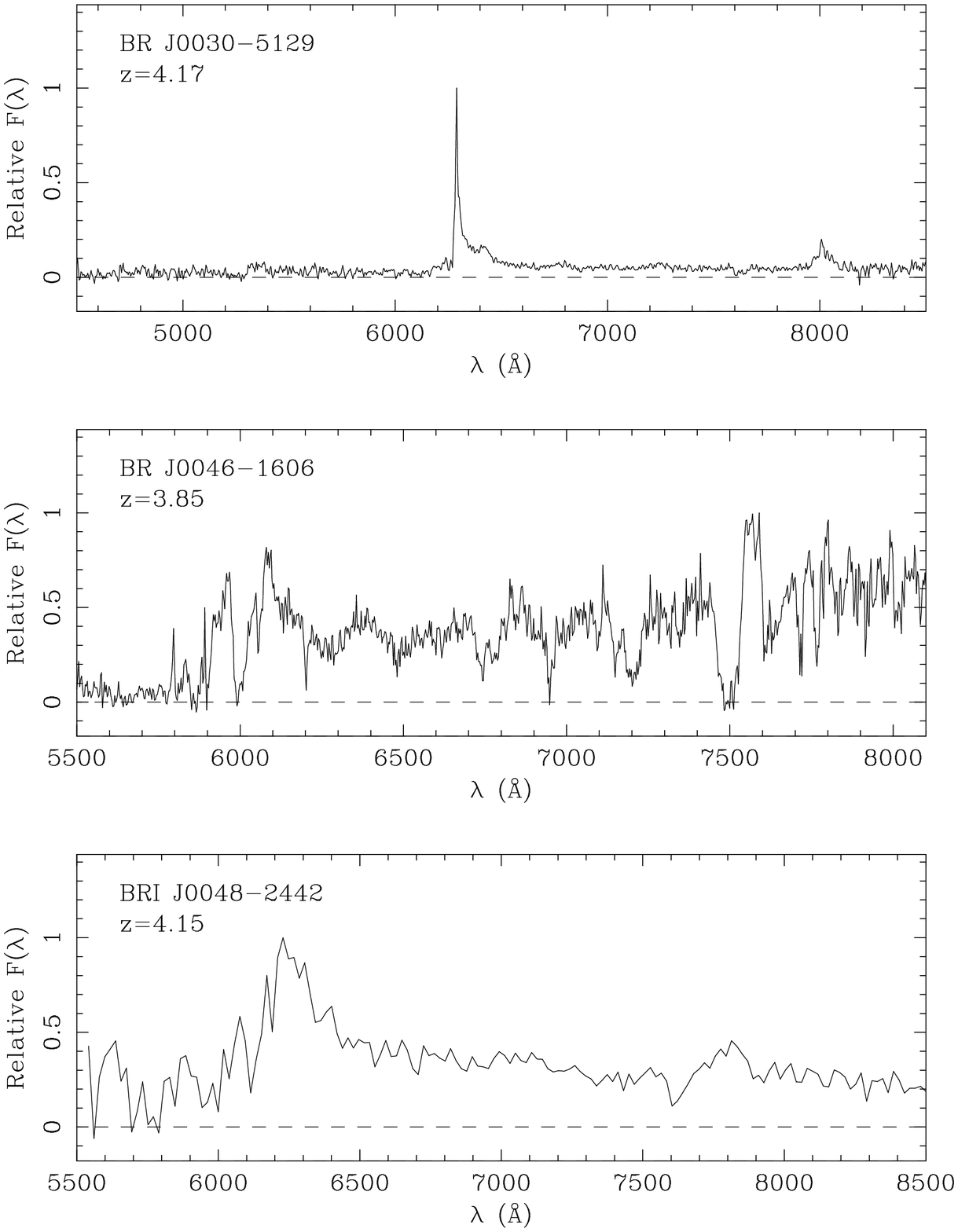,height=10.0cm}}
\end{figure}

\begin{figure}
\centerline{\psfig{figure=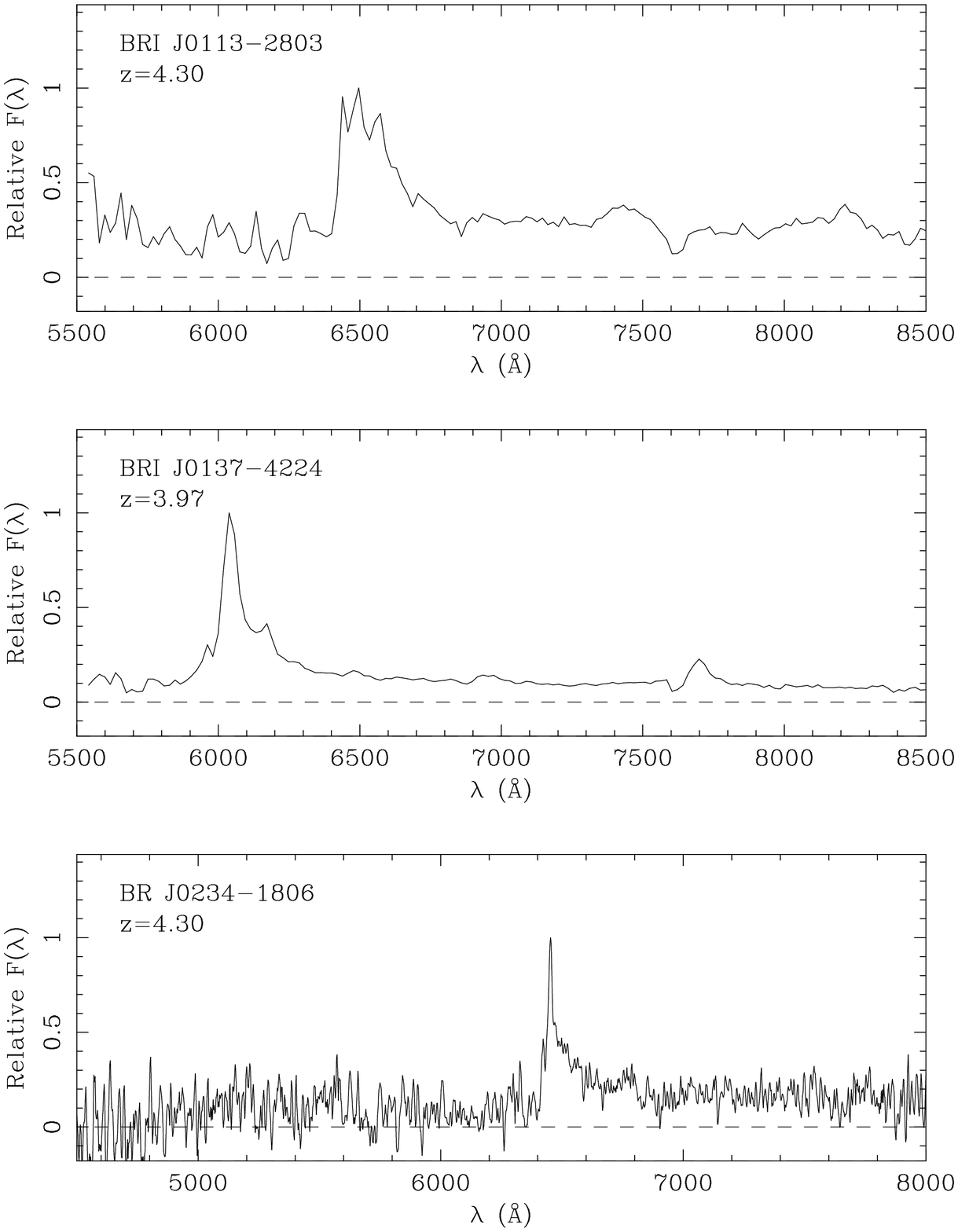,height=10.0cm}}
\centerline{\psfig{figure=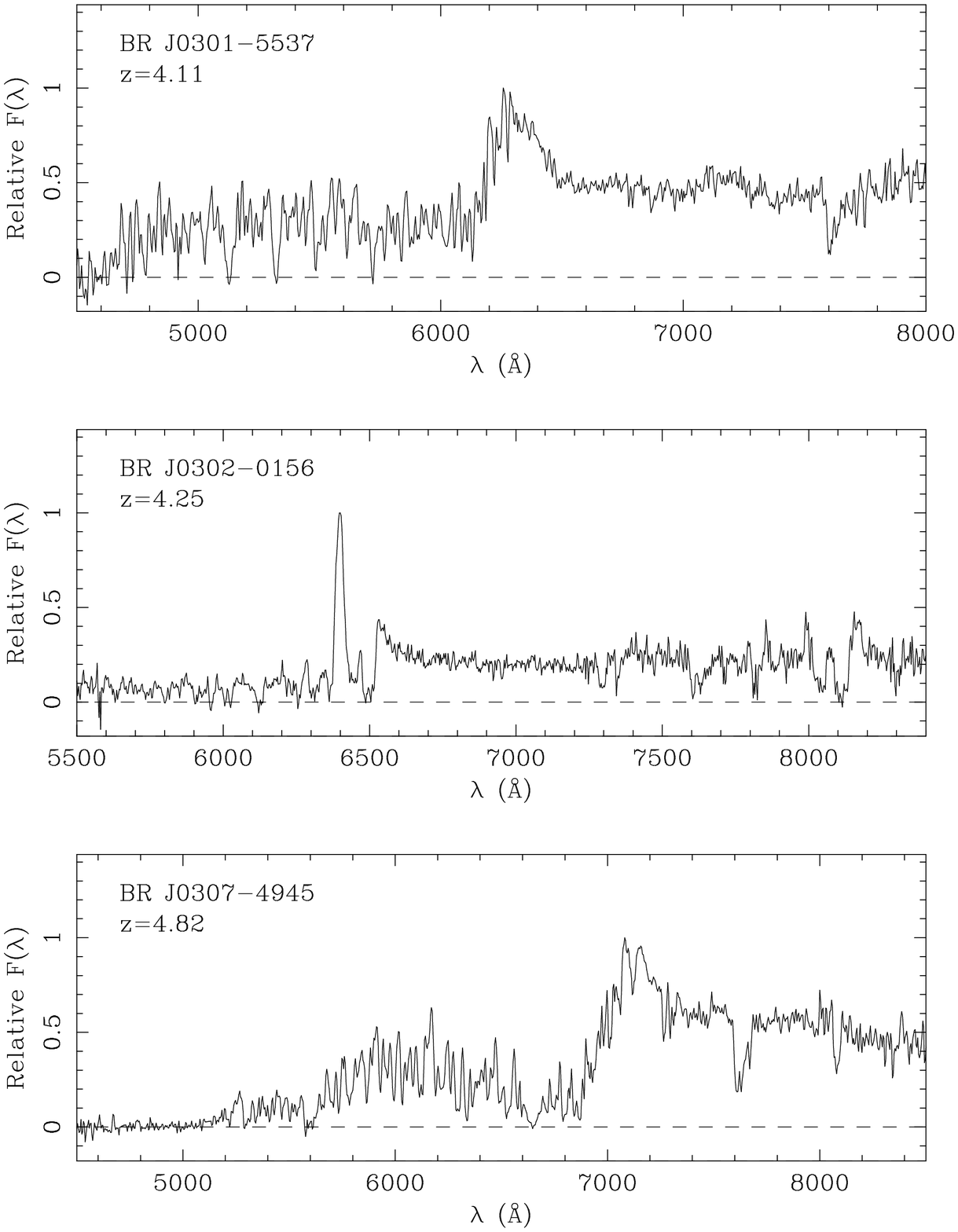,height=10.0cm}}
\caption{The quasar discovery spectra are shown.}
\label{f_spectra}
\end{figure}

\clearpage

\begin{figure}
\centerline{\psfig{figure=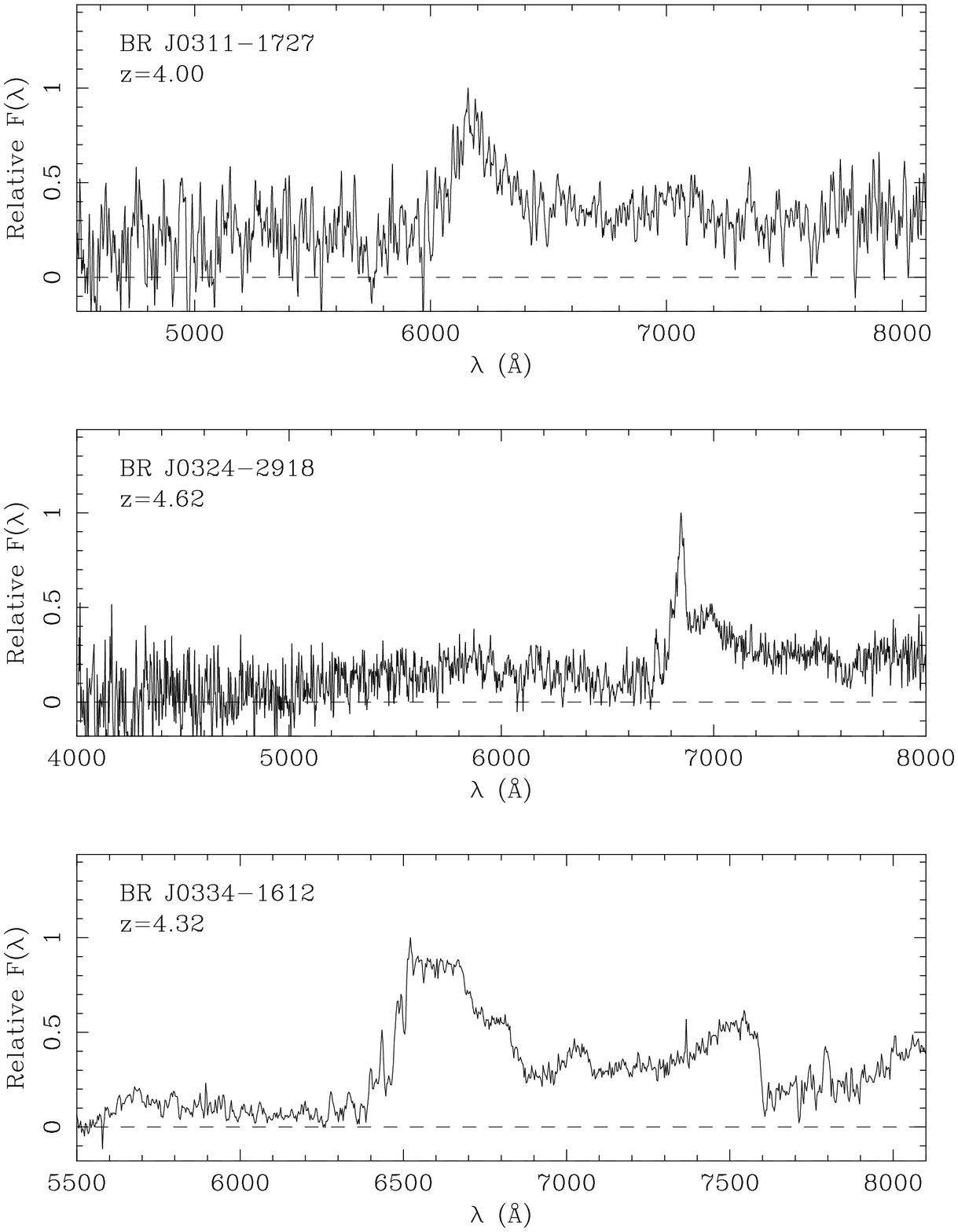,height=10.0cm}}
\centerline{\psfig{figure=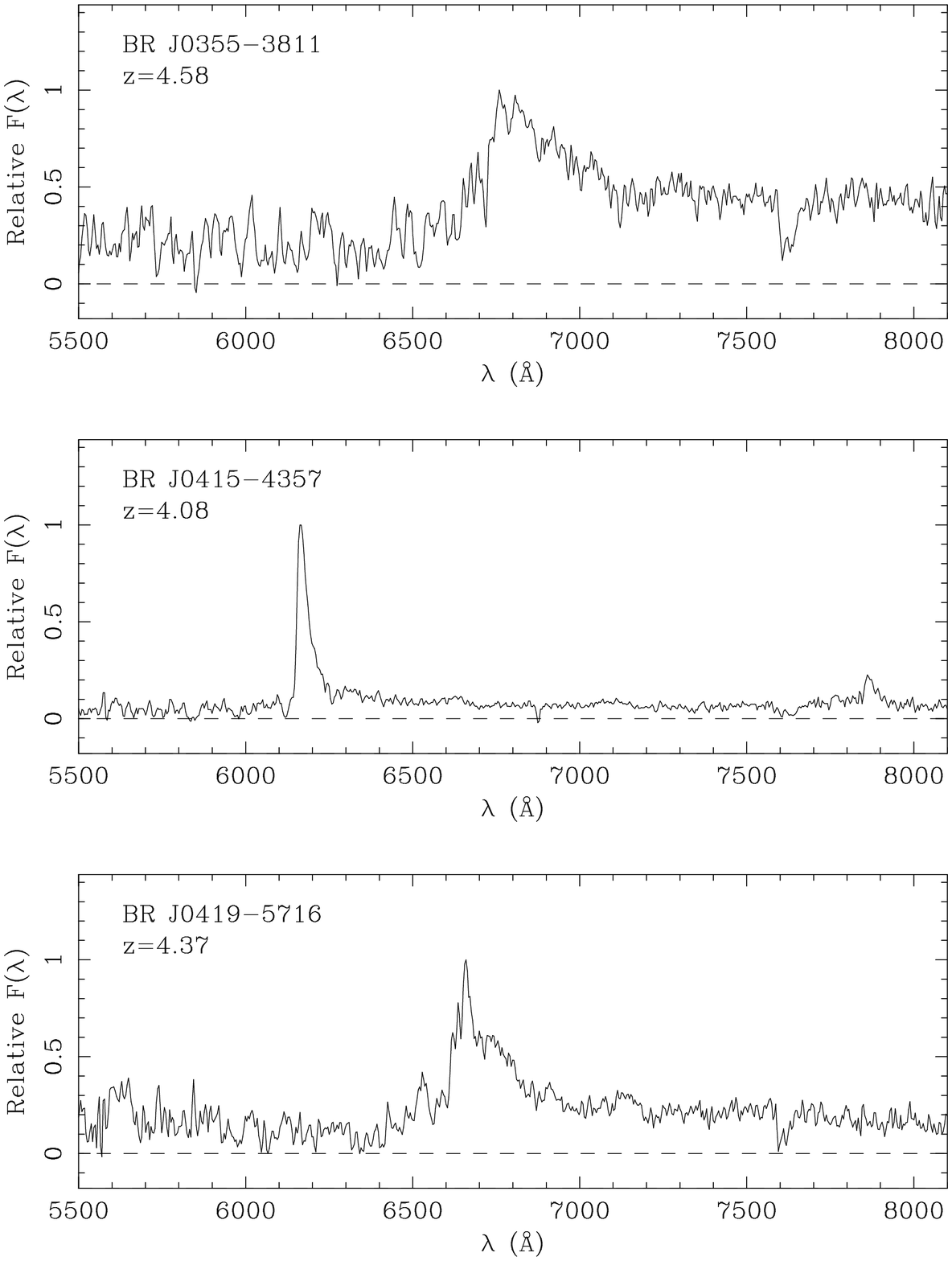,height=10.0cm}}
\end{figure}
 
\begin{figure}
\centerline{\psfig{figure=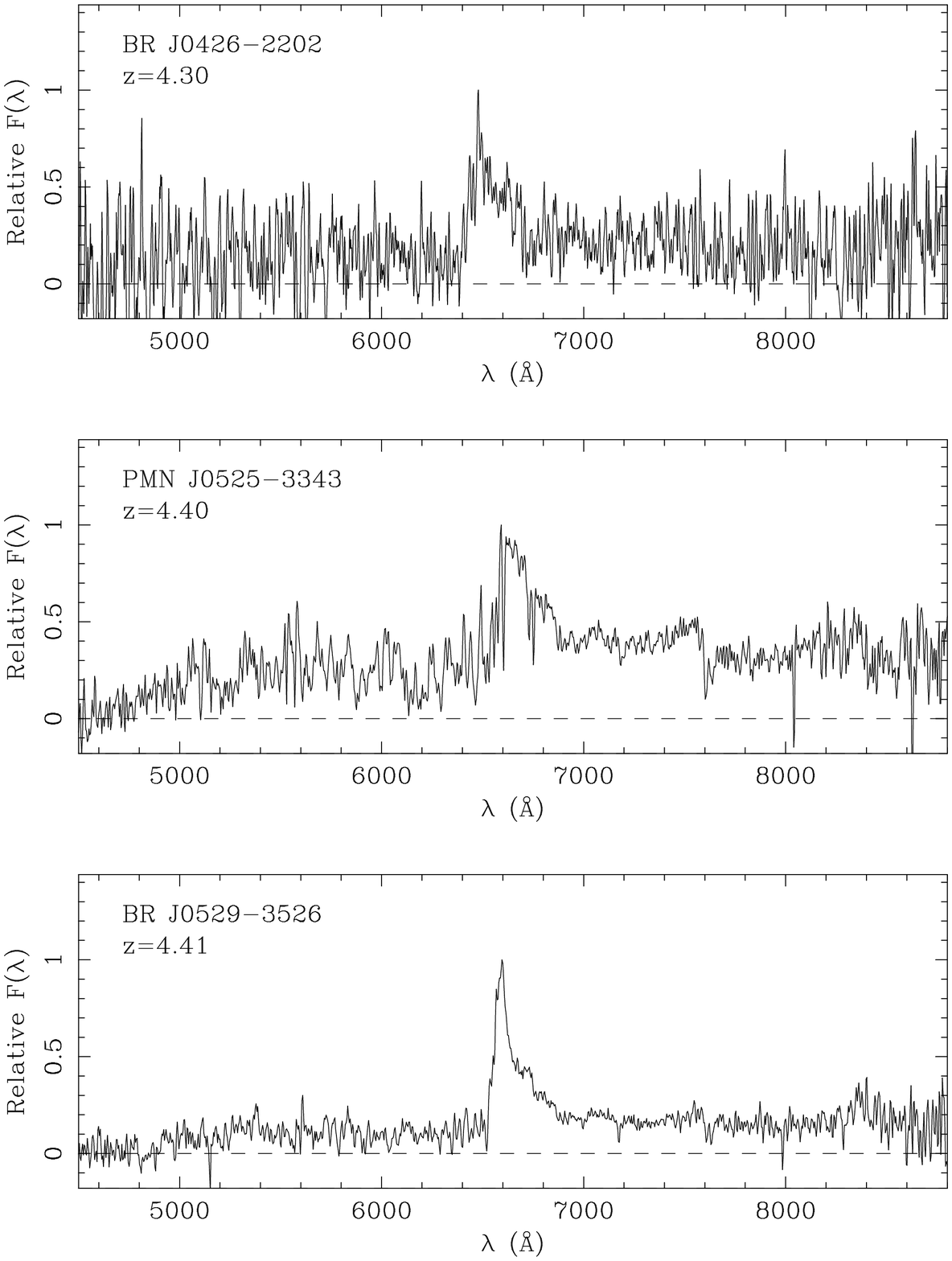,height=10.0cm}}
\centerline{\psfig{figure=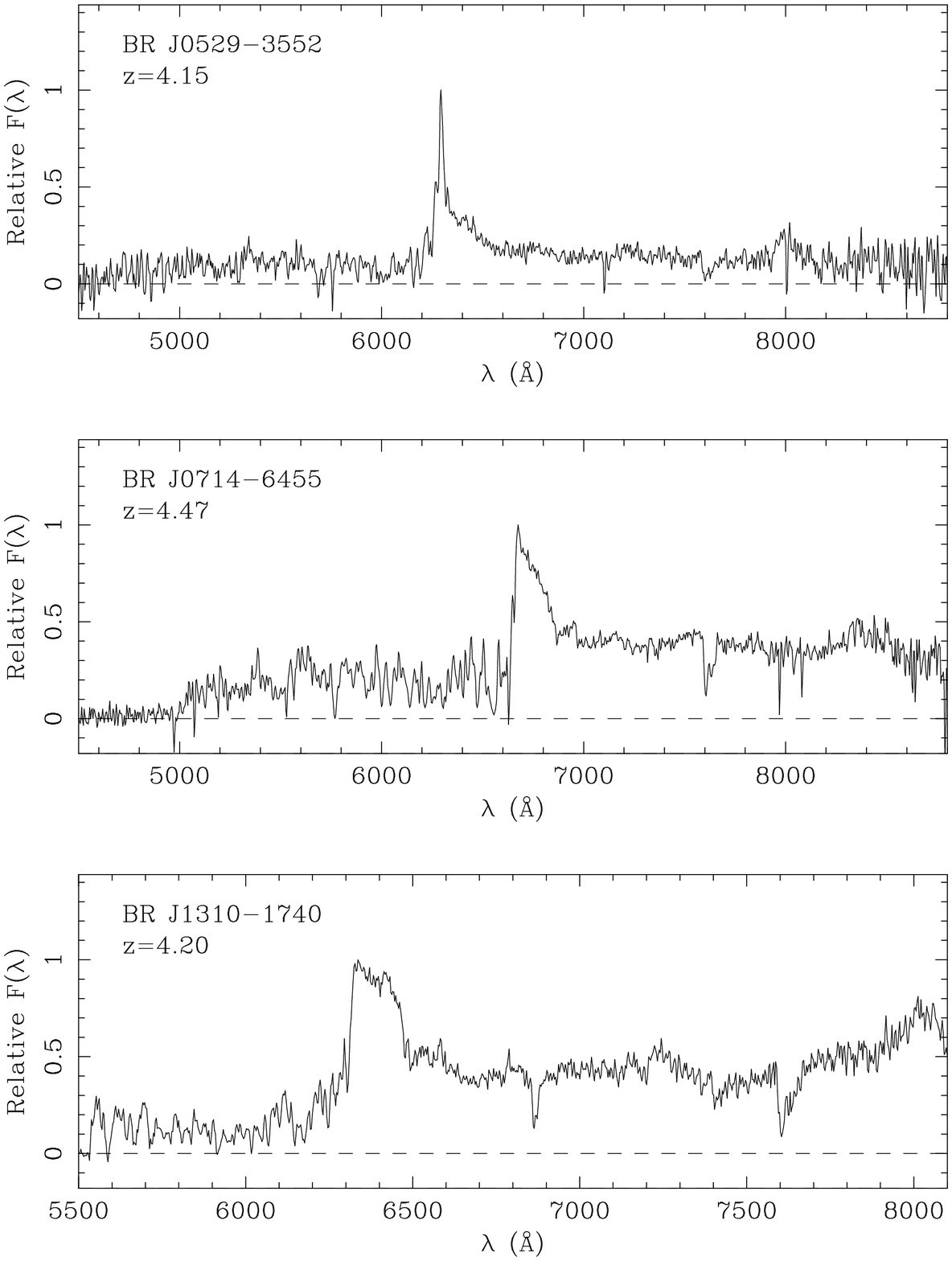,height=10.0cm}}
\contcaption{The quasar discovery spectra are shown.}
\end{figure}
 
\begin{figure}
\centerline{\psfig{figure=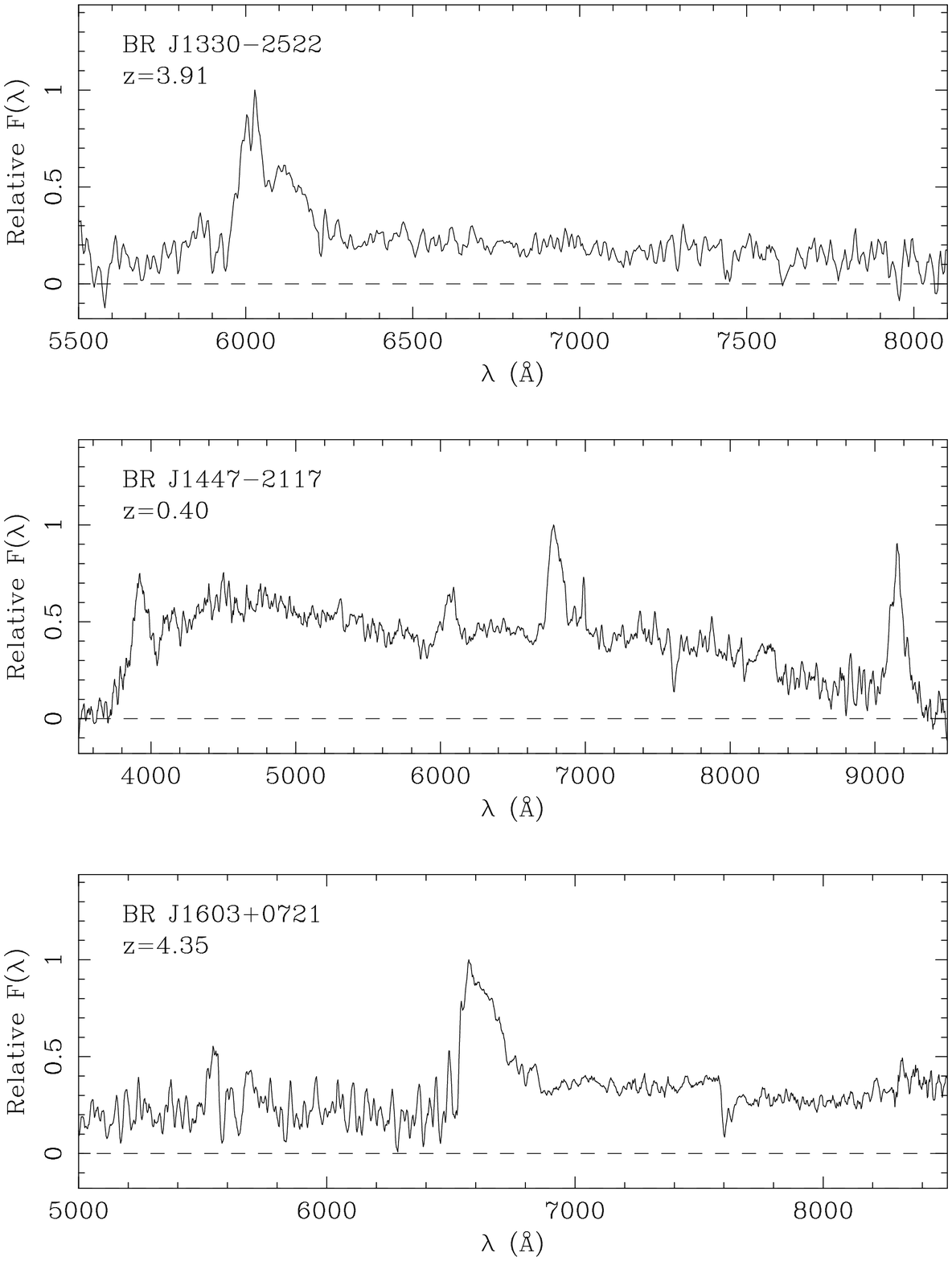,height=10.0cm}}
\centerline{\psfig{figure=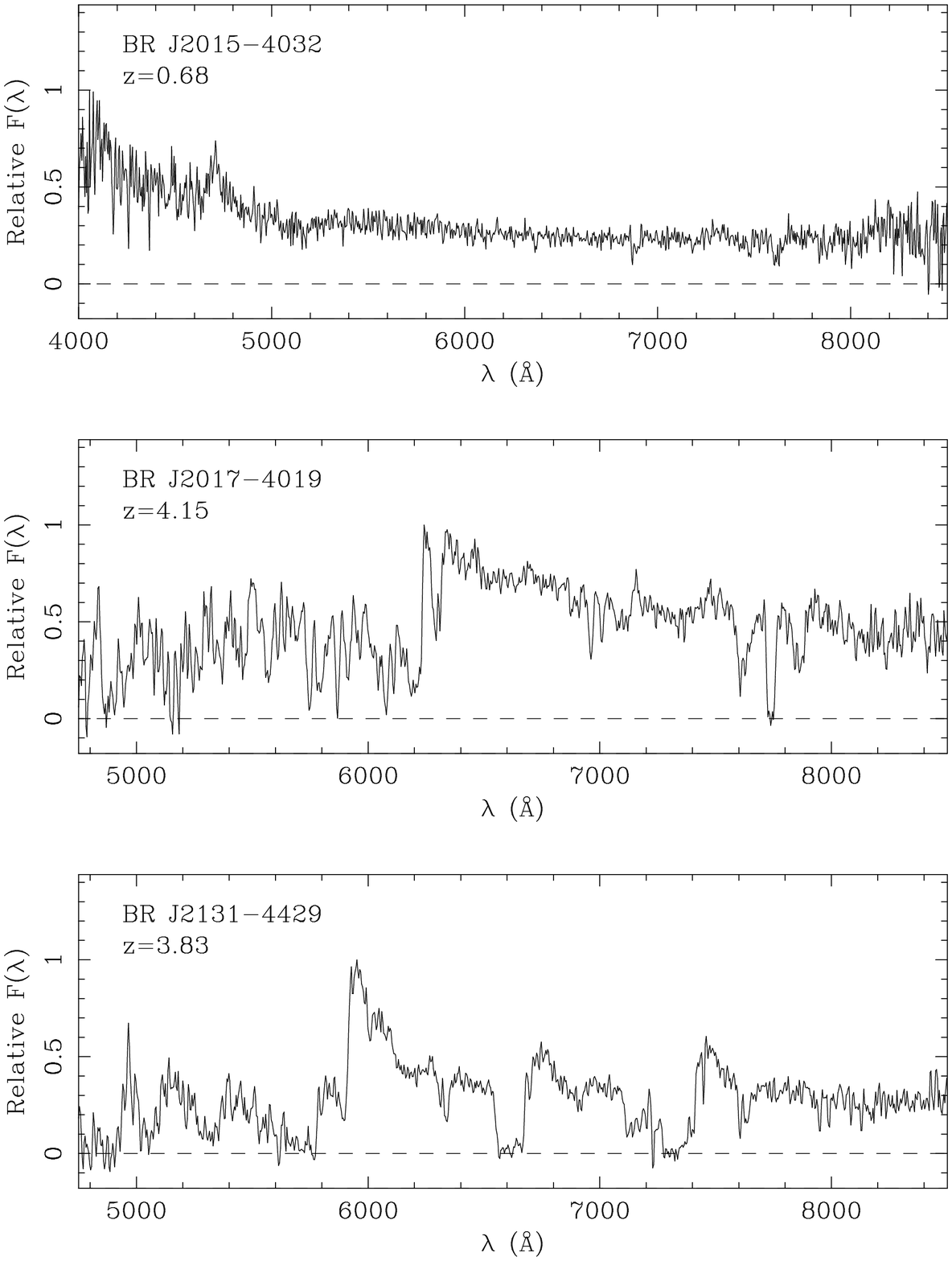,height=10.0cm}}
\end{figure}

\begin{figure}
\centerline{\psfig{figure=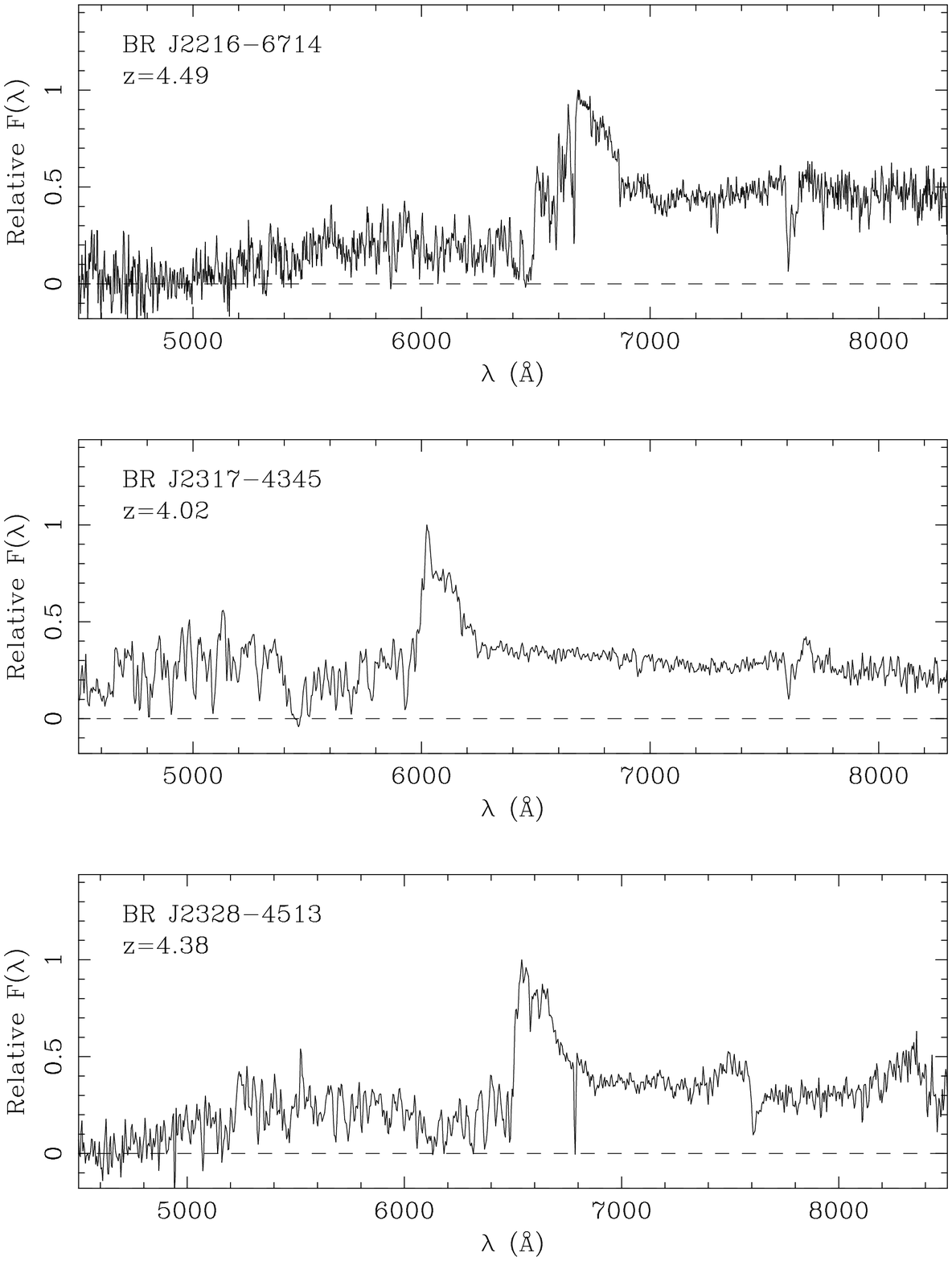,width=8.0cm}}
\centerline{\psfig{figure=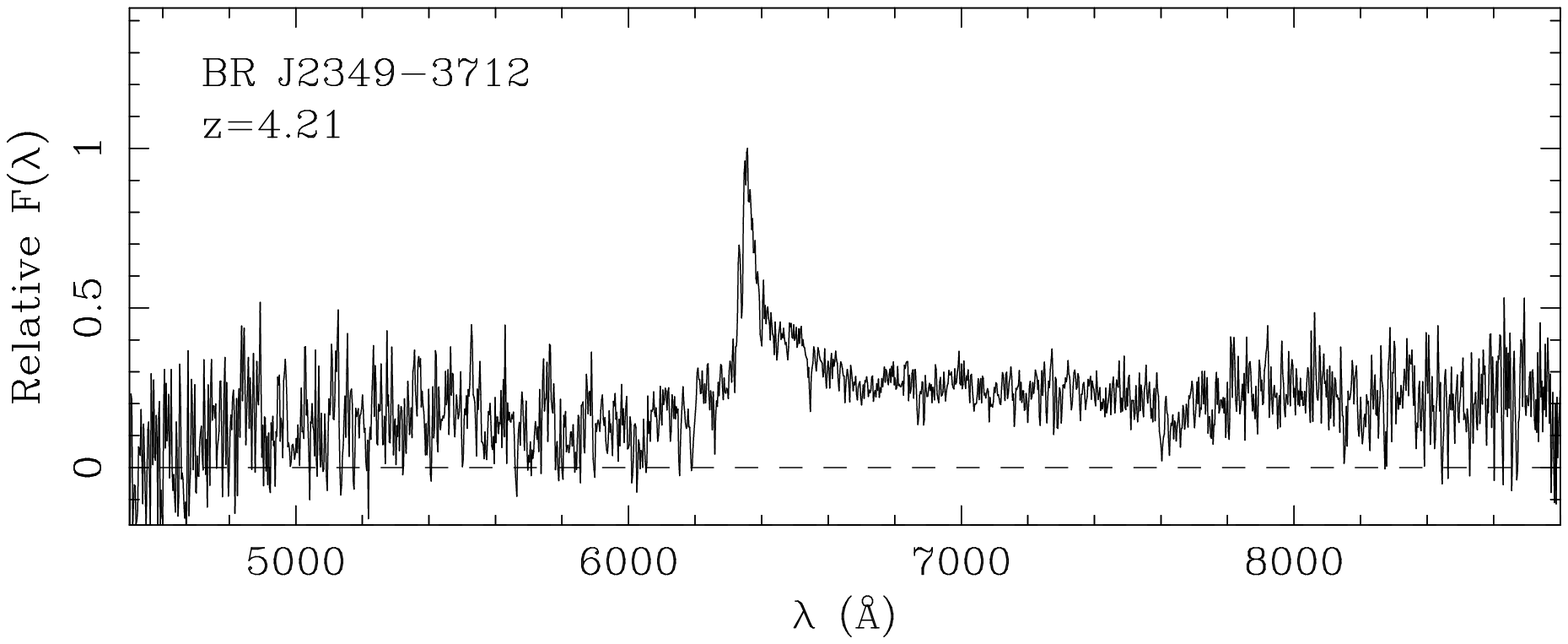,width=8.0cm}}
\contcaption{The quasar discovery spectra are shown.}
\end{figure}
 

\clearpage
\begin{figure*}
\begin{minipage}{115mm}
\centerline{\psfig{figure=z4surveyfig4_1.ps,width=7.0in}}
\caption{The finding charts for the Second APM Colour Survey quasars are
shown. Each is 5 arcminutes on a side, oriented with north up and west to the right.}
\label{f_fc}
\end{minipage}
\end{figure*}

\clearpage
\begin{figure*}
\begin{minipage}{115mm}
\centerline{\psfig{figure=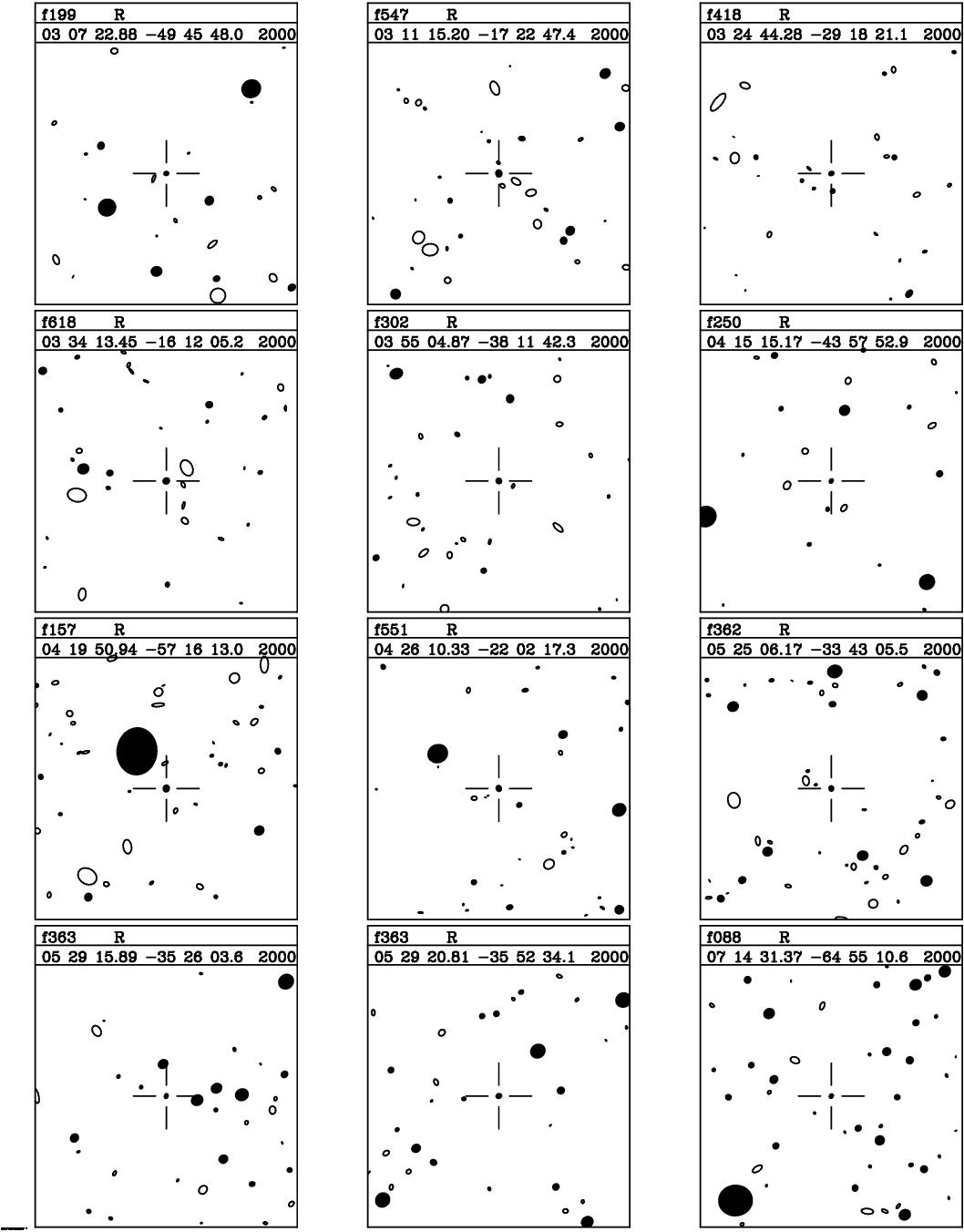,width=7.0in}}
\contcaption{The finding charts for the Second APM Colour Survey quasars are
shown. Each is 5 arcminutes on a side, oriented with north up and west to the right.}
\end{minipage}
\end{figure*}

\clearpage
\begin{figure*}
\begin{minipage}{115mm}
\centerline{\psfig{figure=z4surveyfig4_3.ps,width=7.0in}}
\contcaption{The finding charts for the Second APM Colour Survey quasars are
shown. Each is 5 arcminutes on a side, oriented with north up and west to the right.}
\end{minipage}
\end{figure*}

\clearpage

\begin{figure*}
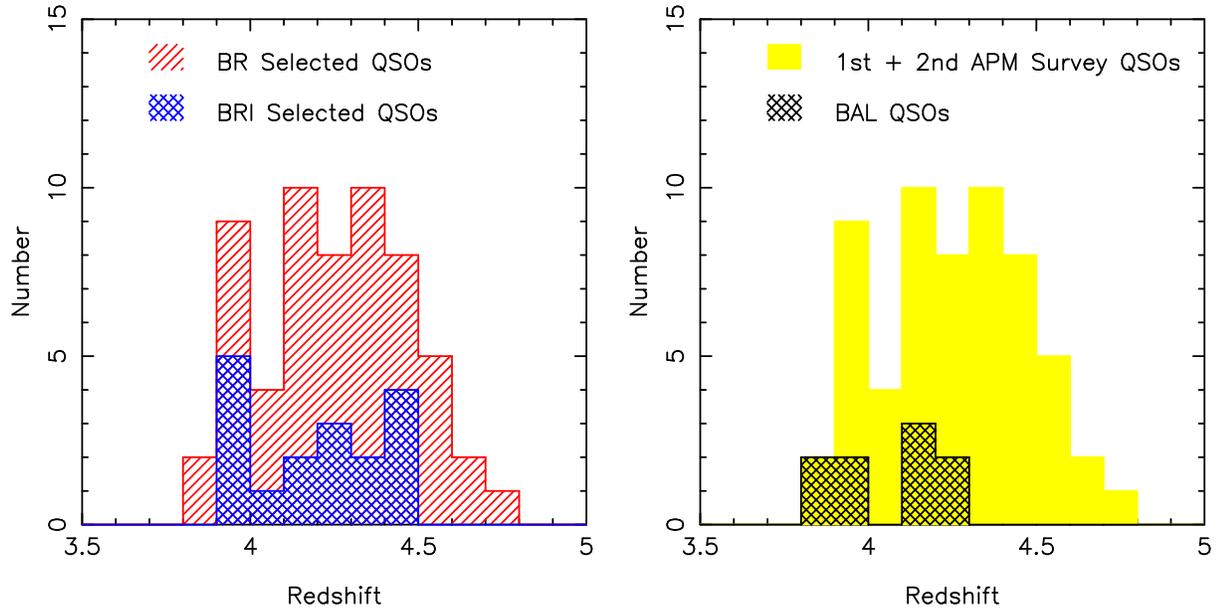

\begin{minipage}{115mm}
\centerline{
\psfig{figure=z4surveyfig5a.ps,height=8.0cm}
\hspace{0.1in}
\psfig{figure=z4surveyfig5b.ps,height=8.0cm}
}
\caption{
These histograms show the combined redshift distribution for the quasars discovered
in the First and Second APM Colour Surveys for z$>$4 Quasars (Storrie-Lombardi
\etal\ 1996; this paper).
The left panel shows the complete histogram with the BRX-selected quasars 
shown with single hatch marks and the BRI-selected quasars shown with the double hatch marks.  
The right panel again shows the combined survey histogram fully shaded, with hatch
marks overlaid on the quasars that exhibit broad absorption lines (BAL) characteristics. }
\label{f_zhist}
\end{minipage}
\end{figure*}

\label{lastpage}

\end{document}